\documentclass[11pt]{article}

\usepackage{amsmath,amssymb,amsthm}
\usepackage{mathtools}
\usepackage{graphicx}
\usepackage{caption, subcaption}
\usepackage{fullpage}
\usepackage{color}
\usepackage[colorlinks=true, citecolor=blue]{hyperref}
\usepackage[usenames,dvipsnames]{xcolor}
\usepackage{tikz}
\usetikzlibrary{knots}

\tikzset{contact/.style={circle, draw=purple, line width=2pt, fill=yellow, inner sep=3pt}}

\usepackage[capitalize,nameinlink,noabbrev]{cleveref}

\usepackage[sc]{mathpazo}
\usepackage[T1]{fontenc}

\newcommand{\x}{\mathbf{x}}
\renewcommand{\u}{\mathbf{u}}

\usepackage[maxbibnames=99,sorting=nyt]{biblatex}
\renewbibmacro{in:}{. In:}
\DeclareFieldFormat[misc]{title}{\mkbibquote{#1}}
\renewcommand*{\bibfont}{\small}
\DeclareFieldFormat[article]{volume}{\mkbibbold{#1}}
\DeclareFieldFormat{url}{\textsc{url}: \href{#1}{#1}}
\DeclareFieldFormat{doi}{\textsc{doi}: \href{http://dx.doi.org/#1}{#1}}
\DeclareFieldFormat{eprint:arxiv}{arXiv:\href{https://arxiv.org/abs/#1}{#1}}

\usepackage{authblk}

\title{Exact solutions of directed walk models of polymeric zipping with pulling in two and three dimensions}
\author{Nicholas R. Beaton\thanks{\href{mailto:nrbeaton@unimelb.edu.au}{nrbeaton@unimelb.edu.au}} }
\author{Aleksander L. Owczarek\thanks{\href{mailto:owczarek@unimelb.edu.au}{owczarek@unimelb.edu.au} \\ The authors gratefully acknowledge support from the Australian Research Council, and in particular grants DE170100186 and DP160103562.}}
\affil{School of Mathematics and Statistics, The University of Melbourne, Australia}

\newcommand{\ol}{\overline}
\newcommand{\olc}{\overline{c}}
\newcommand{\oly}{\overline{y}}
\newcommand{\olu}{\overline{u}}
\newcommand{\olv}{\overline{v}}
\newcommand{\PAF}{P_\mathrm{AF}}
\newcommand{\tAF}{t_\mathrm{AF}}
\newcommand{\PAO}{P_\mathrm{AO}}
\newcommand{\tAO}{t_\mathrm{AO}}
\newcommand{\PSF}{P_\mathrm{SF}}
\newcommand{\tSF}{t_\mathrm{SF}}
\newcommand{\PSO}{P_\mathrm{SO}}
\newcommand{\tSO}{t_\mathrm{SO}}
\newcommand{\WAF}{W_\mathrm{AF}}
\newcommand{\WAO}{W_\mathrm{AO}}
\newcommand{\WSF}{W_\mathrm{SF}}
\newcommand{\WSO}{W_\mathrm{SO}}
\newcommand{\rhoSF}{\rho_\mathrm{SF}}
\newcommand{\sigSF}{\sigma_\mathrm{SF}}

\newcommand{\rhoSO}{\rho_\mathrm{SO}}
\newcommand{\sigSO}{\sigma_\mathrm{SO}}

\newcommand{\rhoAF}{\rho_\mathrm{AF}}
\newcommand{\sigAF}{\sigma_\mathrm{AF}}

\newcommand{\rhoAO}{\rho_\mathrm{AO}}
\newcommand{\sigAO}{\sigma_\mathrm{AO}}

\begin{document}
\maketitle

\begin{abstract}
We provide the exact solution of several variants of simple models of the zipping transition of two bound polymers, such as occurs in DNA/RNA, in two and three dimensions using pairs of directed lattice paths. In three dimensions the solutions are written in terms of complete elliptic integrals. We analyse the phase transition associated with each model giving the scaling of the partition function. We also  extend the models to include a pulling force between one  end of the pair of paths, which competes with the attractive monomer-monomer interactions between the polymers.
\end{abstract}

\section{Introduction}
Experimental techniques able to micro-manipulate single polymers \cite{ashkin1997a-a, strick2001a-a, svoboda1994a-a}
and the connection to modelling DNA denaturation
\cite{essevaz-roulet1997a-a, lubensky2000a-a, lubensky2002a-a, marenduzzo2002a-a, marenduzzo2003a-a, marenduzzo2009a-a, orlandini2001a-a}  have provided the impetus for studying models of polymer adsorption, pulling and zipping. In the pursuit of exact solutions, idealised two-dimensional directed walk models have been constructed to capture the effects of adsorption, where a polymer grafts itself onto a surface at low temperature \cite{bouchaud1989polymer, iliev2012directed, orlandini2004adsorption, privman1988new}; as well as zipping, where two polymers are entwined with one another (again at low temperature) \cite{marenduzzo2001phase, poland1970theory, richard2004poland}. Recently extensions of these models to include multiple effects in two-dimensional exactly solved models of directed walks \cite{owczarek2017a-:a, owczarek2012c-:a, tabbara_exact_2014, tabbara2016a-:a, tabbara2012a-:a} have provided  rich mathematical results that display  key physical characteristics of these polymer systems.  

Here we pursue models of the zipping transition in three dimensions, modelling DNA denaturation, and demonstrate how different variations demonstrate modified, though broadly similar, behaviour. We analyse the scaling behaviour of the associated partition function and the phase transitions that occur. We begin by reviewing and enlarging the range of two-dimensional models solved. The models each contain two directed paths  on either the square or cubic lattice which may share sites.  To these we add an attractive/repulsive potential energy each time they share a such site: this drives the zipping transition where the polymers either come together on average or stay apart.  Our solutions include a pulling force that separates the ends of the walks and so competes with the  zipping interaction. In our models there are three phases which we denote \emph{free}, \emph{zipped} and \emph{unzipped}. It should be noted that without pulling the "zipping" transition is between the free and zipped phases.

\section{Two dimensions}\label{sec:2d}

A directed path $p$ on the square lattice $\mathbb Z^2$ is a sequence of vertices $(p_0,p_1,\dots,p_n)$, with $p_0 = (0,0)$ and $p_i-p_{i-1} \in \{(1,0),(0,1)\}$ for $i=1,\dots,n$. Equivalently, $p$ can be viewed as a sequence of north (N) and east (E) steps.

Let $p$ and $q$ be a (ordered) pair of directed paths of the same length $n$. The pair $p$ and $q$ are \emph{asymmetric} if $x(p_i) \leq x(q_i)$ (equivalently, $y(p_i) \geq y(q_i)$) for all $i$. A pair of paths without the asymmetric restriction are \emph{symmetric} (so asymmetric pairs form a subset of symmetric pairs).
The pair is said to \emph{osculate} if $p_i = q_i \Rightarrow p_{i+1} \neq q_{i+1}$ for all $i$. That is, the two paths never occupy the same edge of the lattice. A pair of paths without the osculating restriction are \emph{friendly} (again, osculating pairs are therefore a subset of friendly pairs).

Let $\mathcal{AO}$ (resp.~$\mathcal{AF}$, $\mathcal{SO}$ and $\mathcal{SF}$) be the set of asymmetric/osculating (resp.~asymmetric/friendly, symmetric/osculating and symmetric/friendly) pairs of paths.

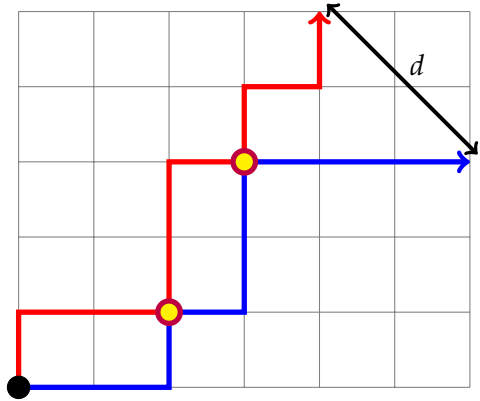
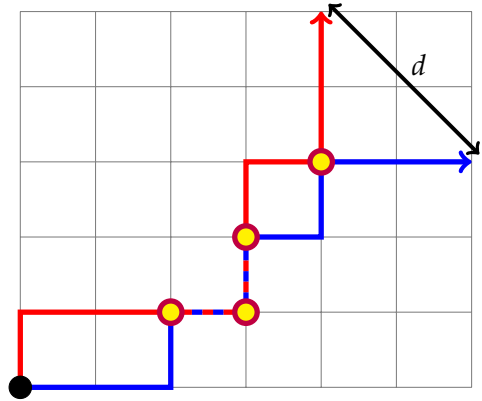
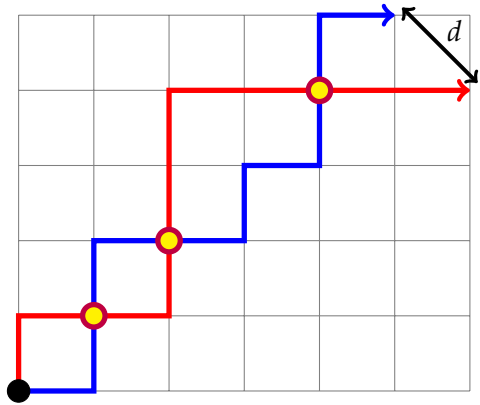
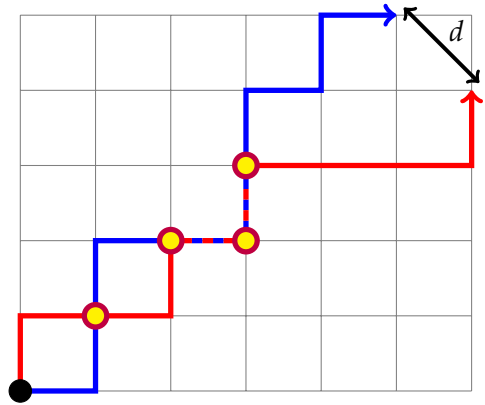
\begin{figure}
    \centering
    \begin{subfigure}{0.46\textwidth}
    \centering
    \begin{tikzpicture}
    \draw [thin, gray] (0,0) grid (6,5);
    \draw [line width=2pt, blue, ->] (0,0) -- (1,0) -- (2,0) -- (2,1) -- (3,1) -- (3,2) -- (3,3) -- (4,3) -- (5,3) -- (6,3);
    \draw [line width=2pt, red, ->] (0,0) -- (0,1) -- (1,1) -- (2,1) -- (2,2) -- (2,3) -- (3,3) -- (3,4) -- (4,4) -- (4,5);
    \node at (0,0) [circle, draw=black, fill=black, inner sep=3pt] {};
    \node at (2,1) [contact] {};
    \node at (3,3) [contact] {};
    \draw [line width=1.5pt, black, <->, xshift=1mm, yshift=1mm] (6,3) -- (4,5);
    \node at (5,4) [xshift=3mm, yshift=3mm] {$d$};
    \end{tikzpicture}
    \caption{$|\phi| = 9$, $v(\phi) = 2$, $d(\phi) = 2$}
    \label{fig:fig_AO}
    \end{subfigure}
    \begin{subfigure}{0.46\textwidth}
    \centering
    \begin{tikzpicture}
    \draw [thin, gray] (0,0) grid (6,5);
    \draw [line width=2pt, blue] (0,0) -- (1,0) -- (2,0) -- (2,1);
    \draw [line width=2pt, blue, dash pattern=on 4pt off 4pt] (2,1) -- (3,1) -- (3,2);
    \draw [line width=2pt, blue, ->] (3,2) -- (4,2) -- (4,3) -- (5,3) -- (6,3);
    \draw [line width=2pt, red] (0,0) -- (0,1) -- (1,1) -- (2,1);
    \draw [line width=2pt, red, dash pattern=on 4pt off 4pt, dash phase=4pt] (2,1) -- (3,1) -- (3,2);
    \draw [line width=2pt, red, ->] (3,2) -- (3,3) -- (4,3) -- (4,4) -- (4,5);
    \node at (0,0) [circle, draw=black, fill=black, inner sep=3pt] {};
    \node at (2,1) [contact] {};
    \node at (3,1) [contact] {};
    \node at (3,2) [contact] {};
    \node at (4,3) [contact] {};
    \draw [line width=1.5pt, black, <->, xshift=1mm, yshift=1mm] (6,3) -- (4,5);
    \node at (5,4) [xshift=3mm, yshift=3mm] {$d$};
    \end{tikzpicture}
    \caption{$|\phi| = 9$, $v(\phi) = 4$, $d(\phi) = 2$}
    \label{fig:fig_AF}
    \end{subfigure}
    
    \vspace{5ex}
    
    \begin{subfigure}{0.46\textwidth}
    \centering
    \begin{tikzpicture}
    \draw [thin, gray] (0,0) grid (6,5);
    \draw [line width=2pt, blue, ->] (0,0) -- (1,0) -- (1,1) -- (1,2) -- (2,2) -- (3,2) -- (3,3) -- (4,3) -- (4,4) -- (4,5) -- (5,5);
    \draw [line width=2pt, red, ->] (0,0) -- (0,1) -- (1,1) -- (2,1) -- (2,2) -- (2,3) -- (2,4) -- (3,4) -- (4,4) -- (5,4) -- (6,4);
    \node at (0,0) [circle, draw=black, fill=black, inner sep=3pt] {};
    \node at (1,1) [contact] {};
    \node at (2,2) [contact] {};
    \node at (4,4) [contact] {};
    \draw [line width=1.5pt, black, <->, xshift=1mm, yshift=1mm] (6,4) -- (5,5);
    \node at (5.5,4.5) [xshift=3mm, yshift=3mm] {$d$};
    \end{tikzpicture}
    \caption{$|\phi| = 10$, $v(\phi) = 3$, $d(\phi) = 1$}
    \label{fig:fig_SO}
    \end{subfigure}
    \begin{subfigure}{0.46\textwidth}
    \centering
    \begin{tikzpicture}
    \draw [thin, gray] (0,0) grid (6,5);
    \draw [line width=2pt, blue] (0,0) -- (1,0) -- (1,1) -- (1,2) -- (2,2);
    \draw [line width=2pt, blue, dash pattern=on 4pt off 4pt] (2,2) -- (3,2) -- (3,3);
    \draw [line width=2pt, blue, ->] (3,3) -- (3,4) -- (4,4) -- (4,5) -- (5,5);
    \draw [line width=2pt, red] (0,0) -- (0,1) -- (1,1) -- (2,1) -- (2,2);
    \draw [line width=2pt, red, dash pattern=on 4pt off 4pt, dash phase=4pt] (2,2) -- (3,2) -- (3,3);
    \draw [line width=2pt, red, ->] (3,3) -- (4,3) -- (5,3) -- (6,3) -- (6,4);
    \node at (0,0) [circle, draw=black, fill=black, inner sep=3pt] {};
    \node at (1,1) [contact] {};
    \node at (2,2) [contact] {};
    \node at (3,2) [contact] {};
    \node at (3,3) [contact] {};
    \draw [line width=1.5pt, black, <->, xshift=1mm, yshift=1mm] (6,4) -- (5,5);
    \node at (5.5,4.5) [xshift=3mm, yshift=3mm] {$d$};
    \end{tikzpicture}
    \caption{$|\phi| = 10$, $v(\phi) = 4$, $d(\phi) = 1$}
    \label{fig:fig_SF}
    \end{subfigure}
    \caption{The four types of two-dimensional pairs of paths: (a) asymmetric/osculating, (b) asymmetric/friendly, (c) symmetric/osculating, and (d) symmetric/friendly.}
    \label{fig:2D_examples}
\end{figure}

We define the following three statistics on pairs of paths $\phi=(p,q)$ of length $n$:
\begin{itemize}
\item $|\phi| = n$;
\item $v(\phi) = |\{i > 0 : p_i = q_i\}|$, that is, the number of shared vertices (excluding the origin);
\item $d(\phi) = \frac{1}{\sqrt{2}}\|q_n - p_n\|$, that is, the (scaled) separation of the endpoints.
\end{itemize}
Note that $d$ is equivalent to the minimum number of steps that $p$ and $q$ must take in order to come together. See \cref{fig:2D_examples} for examples.

For each of the four sets $\mathcal X$, define the partition functions
\begin{equation}
X_n(c,y) = \sum_{\substack{\phi\in\mathcal{X} \\ |\phi| = n}}c^{v(\phi)} y^{d(\phi)}.
\end{equation}
The variables $c$ and $y$ are Boltzmann weights, and can be interpreted as $c=e^{\alpha/kT}$ and $y=e^f$, where $\alpha$ is the energy associated with a contact between the two polymers, $T$ is absolute temperature, $k$ is Boltzmann's constant and $f$ is a force applied to the endpoints of the polymers, pulling them apart when $f>0$ and together when $f<0$.

The free energy of the system is
\begin{equation}
\psi_\mathrm{X}(c,y) = \lim_{n\to\infty} \frac1n \log X_n(c,y).
\end{equation}

It will also be useful to define the generating functions
\begin{equation}
P_\mathrm{X}(t;c,y) = \sum_n X_n(c,y)t^n = \sum_{\phi\in\mathcal{X}}t^{|\phi|} c^{v(\phi)} y^{d(\phi)}.
\end{equation}
These will be viewed as power series in $t$ with coefficients in $\mathbb Z[c,y]$. Note that if $t_\mathrm{X}(c,y)$ is the radius of convergence of this series, then
\begin{equation}
\psi_\mathrm{X}(c,y) = -\log t_\mathrm{X}(c,y).
\end{equation}
For brevity we will often write $P_\mathrm{X}(y)$ instead of $P_\mathrm{X}(t;c,y)$.

\subsection{Asymmetric and friendly pairs of paths}

The four two-dimensional models can all be solved with a now-classical tool called the \emph{kernel method} \cite{prodinger_kernel_2004}. We will give the details for asymmetric/friendly pairs.

Pairs of paths are iteratively grown one pair of steps at a time. Initially, a pair consists only of a single vertex. After this, each of the two paths $(p,q)$ can step N or E, subject to the asymmetric constraint that $p$ cannot step to the right of $q$. When $p$ steps N and $q$ steps E, $v$ increases by 1; when $p$ steps E and $q$ steps N, $v$ decreases by 1; and in the other two cases $v$ does not change. In addition, when the pair step to a shared vertex, $c$ increases by 1.

This all gives the functional equation
\begin{equation}\label{eqn:PAF_fe}
\PAF(y) = 1 + t(2+y+\oly)\PAF(y) - t\oly\PAF(0) + 2t(c-1)\PAF(0) + t(c-1)[y^1]\PAF(y),
\end{equation}
where $\oly = \frac1y$ and $[y^1]$ is the linear operator which extracts the coefficient of $y^1$ from each term of a power series.

We can eliminate the $[y^1]\PAF(y)$ term by considering those pairs which end together:
\begin{equation}\label{eqn:PAF_bdy}
\PAF(0) = 1 + 2tc\PAF(0) + tc[y^1]\PAF(y).
\end{equation}
Combining~\eqref{eqn:PAF_fe} and~\eqref{eqn:PAF_bdy},
\begin{equation}\label{eqn:PAF_cancelled}
K(y)\PAF(y) = \olc + (1-\olc -t\oly)\PAF(0),
\end{equation}
where $K(y) \equiv K(t;y) = 1-t(2+y+\oly)$ and $\olc = \frac1c$.

The kernel $K(y)$ has two roots in $y$; one of them,
\begin{equation}
Y \equiv Y(t) = \frac{1-2t-\sqrt{1-4t}}{2t},
\end{equation}
has a power series expansion around $t=0$. Substituting into~\eqref{eqn:PAF_cancelled} cancels the left side, yielding
\begin{equation}
\PAF(0) = \frac{Y}{tc+(1-c)Y}.
\end{equation}
Substituting this into~\eqref{eqn:PAF_cancelled} then gives the overall solution
\begin{align}
\PAF(y) &= \frac{t(y-Y)}{yK(y)(tc+(1-c)Y)} \\
&= \frac{2\left(1-2t(1+t)+\sqrt{1-4t}\right)}{\left(1+2t-2t(2+t)c+\sqrt{1-4t}\right)\left(1-2t(1+y)+\sqrt{1-4t}\right)}.
\end{align}

For given $c$ and $y$, the radius of convergence of $\PAF(c,y)$ is given by the absolute value of the dominant singularity, ie.~the closest point of non-analyticity to the origin. In $\PAF(c,y)$, there are three possible sources of singularities -- the branch point of the square root in $Y$, roots of $K(y)$, and roots of $tc+(1-c)Y$.

These singularities all play a part in the asymptotics of the model, and their locations are respectively
\begin{equation}
\frac14, \qquad \frac{y}{(1+y)^2}, \qquad\text{and}\qquad \frac{1-c\pm\sqrt{c(c-1)}}{c}.
\end{equation}
By examining how these functions vary with $c$ and $y$, it is straightforward to determine that the dominant singularity of the model is
\begin{equation}
\tAF(c,y) = \begin{cases} \textstyle \frac14 & \text{ if } y\leq 1\text{ and }c\leq \frac43 \\
\textstyle \frac{y}{(1+y)^2} & \text{ if } y\geq\max\{1,f(c)\} \\
\frac{1-c+\sqrt{c(c-1)}}{c} & \text{ if } c \geq \frac43 \text{ and } y \leq f(c) \end{cases}
\end{equation}
where
\[f(c) = c-1+\sqrt{c(c-1)}.\]
The three regions correspond respectively to the free, ballistic and zipped phases. The free-zipped boundary is at $c=\frac43$, the free-ballistic boundary is at $y=1$, and the zipped-ballistic boundary is at $y=f(c)$.

\subsection{Asymmetric and osculating pairs of paths}

A similar application of the kernel method yields
\begin{align}
\PAO(t;c,y) &= \frac{t(y-Y)}{yK(y)(tc+(1-c+2tc)Y)} \\
&= \frac{2\left(1-2t(2-t)+(1-2t)\sqrt{1-4t}\right)}{\left(1-2t-2t^2c+\sqrt{1-4t}\right)\left(1-2t(1+y)+\sqrt{1-4t}\right)}.
\end{align}
The dominant singularity is
\begin{equation}
\tAO(c,y) = \begin{cases}\textstyle \frac14 & \text{ if } y\leq1 \text{ and } c\leq4 \\
\textstyle \frac{y}{(1+y)^2} & \text{ if } y\geq\max\{1,g(c)\} \\
\textstyle \frac{\sqrt{c}-1}{c} & \text{ if } c \geq 4 \text{ and }y\leq g(c) \end{cases}
\end{equation}
where
\[g(c) = \sqrt{c}-1.\]
The three regions correspond respectively to free, ballistic and zipped phases. The free-zipped boundary is at $c=4$, the free-ballistic boundary is at $y=1$, and the zipped-ballistic boundary is at $y=g(c)$.

\subsection{Symmetric and friendly pairs of paths}

For symmetric pairs the paths can cross, but we will weight the endpoint separation regardless of which path is above or below.

Another application of the kernel method gives the generating function as
\begin{align}
\PSF(t;c,y) &= \frac{t(y-Y)(1+yY)}{yK(y)(2tc+(1-2c+2tc)Y)} \\
&= \frac{t(1-y^2)-y\sqrt{1-4t}}{\left(1-c+c\sqrt{1-4t}\right)\left(t(1+y)^2-y\right)}
\end{align}

The dominant singularity is then
\begin{equation}
\tSF(c,y) = \begin{cases} \textstyle \frac14 & \text{ if } y\leq 1 \text{ and } c\leq 1 \\
\textstyle \frac{y}{(1+y)^2} & \text{ if } y\geq\max\{1,h(c)\} \\
\textstyle \frac{2c-1}{4c^2} & \text{ if } c\geq 1 \text{ and } y \leq h(c) \end{cases}
\end{equation}
where
\[h(c) = 2c-1.\]

\subsection{Symmetric and osculating pairs of paths}

The kernel method gives the generating function as
\begin{align}
\PSO(t;c,y) &= \frac{t(y-Y)(1+yY)}{yK(y)(2tc + (1-2c+4tc)Y)} \\
&= \frac{t(1-y^2)-y\sqrt{1-4t}}{\left(1-c+2tc+c\sqrt{1-4t}\right)\left(t(1+y)^2-y\right)}
\end{align}
The dominant singularity is
\begin{equation}
\tSO(c,y) = \begin{cases} \frac14 & \text{ if } y\leq 1 \text{ and } c\leq 2 \\
\frac{y}{(1+y)^2} & \text{ if } y \geq \max\{1,m(c)\} \\
\frac{\sqrt{2c}-1}{2c} & \text{ if } c\geq2 \text{ and } y\leq m(c) \end{cases}
\end{equation}
where
\[m(c) = \sqrt{2c}-1.\]
The free-zipped boundary is $c=2$, the free-ballistic boundary is $y=1$, and the zipped-ballistic boundary is $y=m(c)$. Note that $\tSO(c,y) = \tAO(2c,y)$.

\subsection{Comparing the four models}

\begin{figure}
\centering
\begin{tikzpicture}
\node at (0,0) [anchor=south west] {\includegraphics[width=0.7\textwidth]{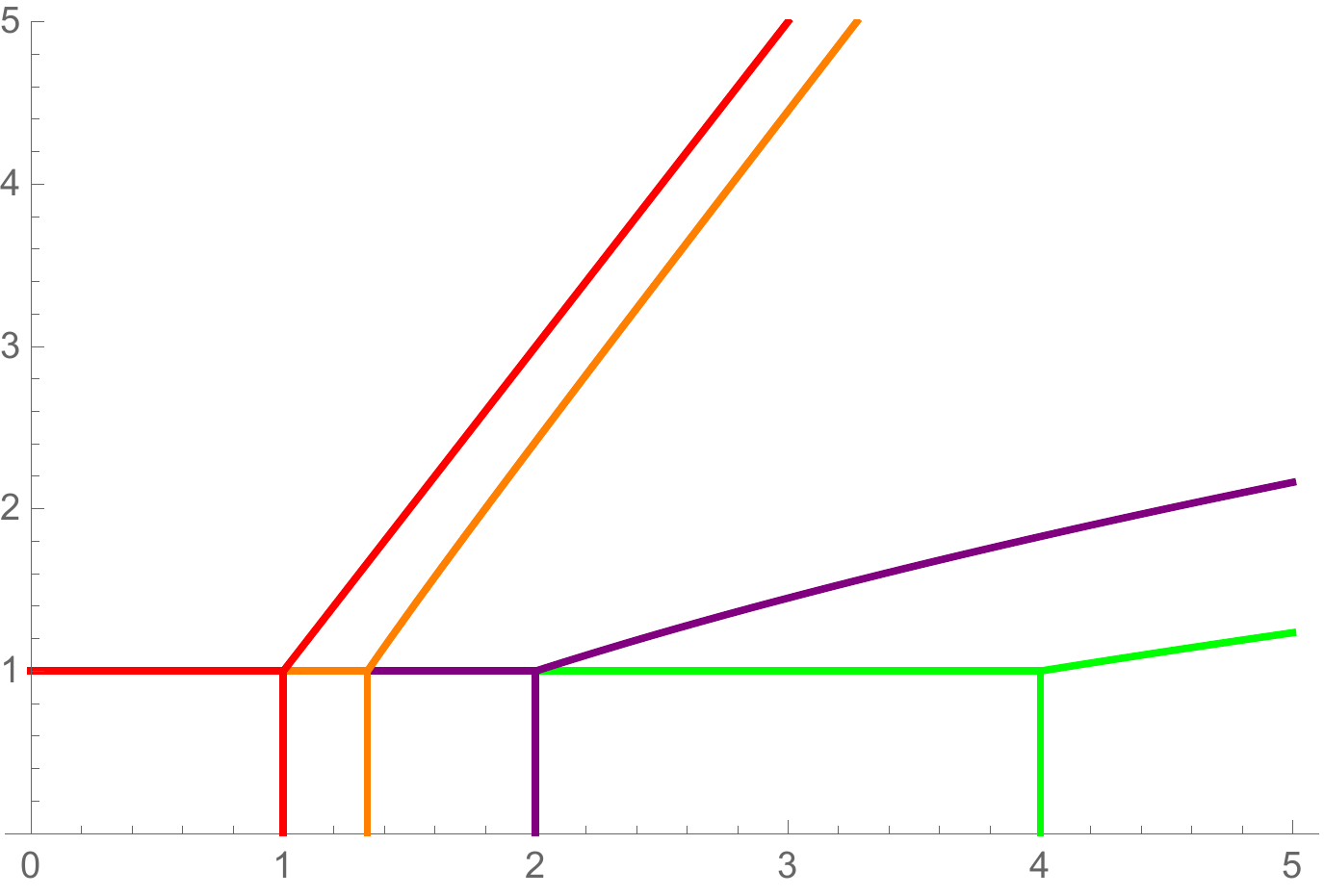}};
\node at (4.3,4.9) {SF};
\node at (5.8,4.5) {AF};
\node at (10.2,3.85) {SO};
\node at (10.2,2.6) {AO};
\node at (6.2,0) {$c$};
\node at (0,4.3) {$y$};
\end{tikzpicture}
\caption{The phase boundaries for the four two-dimensional models. The vertical lines are the boundaries between the free and zipped phases; the horizontal lines are the boundaries between free and unzipped; and the sloping curves are the unzipped-zipped boundaries.}
\label{fig:2d_phase_boundaries}
\end{figure}
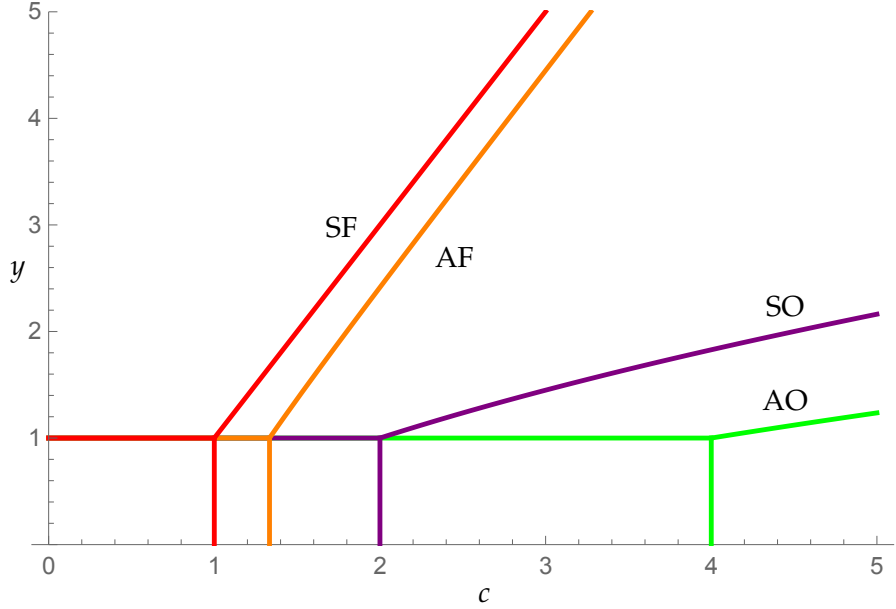

See \cref{fig:2d_phase_boundaries} for a plot of the four different phase boundaries.

For fixed $y$, the zipping transitions occur with increasing $c$ in the order
\[\mathrm{SF} < \mathrm{AF} < \mathrm{SO} < \mathrm{AO}.\]
By looking at the entropic loss involved in a contact, this makes sense: SF loses no entropy at a contact, AF loses one of its four ``choices'', SO loses two of four choices, and AO loses three of four.

In all four cases, the free-zipped and free-unzipped phase boundaries are second-order, while the zipped-unzipped phase boundaries are first-order.

\section{Three dimensions}

In three dimensions we again take a pair $\phi=(p,q)$ of directed paths (ie.~paths which step in the positive $x$, $y$ or $z$ directions) which start at the origin. However, unlike in two dimensions, there is no longer a sensible notion of the paths ``crossing''. In order to generalize the notion of symmetric and asymmetric pairs to three dimensions, we will say that the pair is \emph{asymmetric} if they satisfy the following: if $p_i = q_i$, then
\begin{itemize}
\item $p_{i+1}-p_i = (1,0,0) \Rightarrow q_{i+1}-q_i \neq (0,0,1)$
\item $p_{i+1}-p_i = (0,1,0) \Rightarrow q_{i+1}-q_i \neq (1,0,0)$
\item $p_{i+1}-p_i = (0,0,1) \Rightarrow q_{i+1}-q_i \neq (0,1,0)$.
\end{itemize}
That is, if $p$ and $q$ share vertex $i$ and $p$'s next step is $+x$ (resp.~$+y, +z$), then $q$'s next step is \emph{not} $+z$ (resp.~$+x,+y$). \emph{Symmetric} pairs are not restricted in this way.

Osculating and friendly paths are defined as for two dimensions (friendly paths may share edges, osculating paths may not).

As in 2D, we let $\mathcal{AO}$ (resp.~$\mathcal{AF}$, $\mathcal{SO}$ and $\mathcal{SF}$) be the set of asymmetric/osculating (resp.~asymmetric/friendly, symmetric/osculating and symmetric/friendly) pairs of paths.

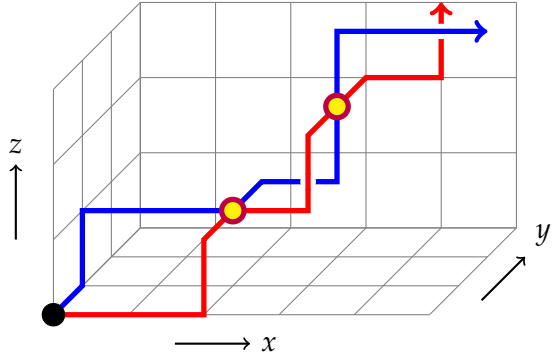
\begin{figure}
    \centering
    \begin{tikzpicture}[rotate around x=270]
    \draw [thin, gray] (0,0,0) -- (5,0,0) (0,1,0) -- (5,1,0) (0,2,0) -- (5,2,0) (0,3,0) -- (5,3,0);
    \draw [thin, gray] (0,0,0) -- (0,3,0) (1,0,0) -- (1,3,0) (2,0,0) -- (2,3,0) (3,0,0) -- (3,3,0) (4,0,0) -- (4,3,0) (5,0,0) -- (5,3,0);
    \draw [thin, gray] (0,0,0) -- (0,0,3) (0,1,0) -- (0,1,3) (0,2,0) -- (0,2,3) (0,3,0) -- (0,3,3);
    \draw [thin, gray] (0,0,0) -- (0,3,0) (0,0,1) -- (0,3,1) (0,0,2) -- (0,3,2) (0,0,3) -- (0,3,3);
    \draw [thin, gray] (0,3,0) -- (5,3,0) (0,3,1) -- (5,3,1) (0,3,2) -- (5,3,2) (0,3,3) -- (5,3,3);
    \draw [thin, gray] (0,3,0) -- (0,3,3) (1,3,0) -- (1,3,3) (2,3,0) -- (2,3,3) (3,3,0) -- (3,3,3) (4,3,0) -- (4,3,3) (5,3,0) -- (5,3,3);
    \begin{knot}[consider self intersections=true, clip width=3, clip radius=3pt, end tolerance=1pt]
    \strand [line width=2pt, red, ->] (0,0,0) -- (1,0,0) -- (2,0,0) -- (2,0,1) -- (2,1,1) -- (3,1,1) -- (3,1,2) -- (3,2,2) -- (3,3,2) -- (4,3,2) -- (4,3,3);
    \strand [line width=2pt, blue, ->] (0,0,0) -- (0,1,0) -- (0,1,1) -- (1,1,1) -- (2,1,1) -- (2,2,1) -- (3,2,1) -- (3,2,2) -- (3,2,3) -- (4,2,3) -- (5,2,3);
    \flipcrossings{2}
    \end{knot}
    \node at (0,0,0) [circle, draw=black, fill=black, inner sep=3pt] {};
    \node at (2,1,1) [contact] {};
    \node at (3,2,2) [contact] {};
    \draw [thick, ->] (2,-1,0) -- (3,-1,0) node [right] {$x$};
    \draw [thick, ->] (5.5,0.5,0) -- (5.5,2,0) node [above right] {$y$};
    \draw [thick, ->] (-0.5,0,1) -- (-0.5,0,2) node [above] {$z$};
    \end{tikzpicture}
    \caption{An asymmetric/osculating pair of paths $\phi$ with $|\phi|=10$, $v(\phi) = 2$, $d_x(\phi) = -1$, $d_y(\phi) = 1$ and $d_z(\phi) = 0$. The red path is $p$ and the blue path is $q$.}
    \label{fig:3d_AO}
\end{figure}

If $\phi = (p,q)$ is a pair of paths, we again let $|\phi|$ be the length of $p$ and $q$ and $v(\phi)$ be the number of shared vertices, excluding the origin. However, the statistic $d(\phi)$ must be defined slightly differently in three dimensions. We will postpone its definition for now. We instead introduce two new measurements: $d_x(\phi) = x(p_n)-x(q_n)$ and $d_y(\phi) = y(p_n) - y(q_n)$. Note that we can also define $d_z(\phi) = z(p_n) - z(q_n)$, but
\begin{equation}
d_z(\phi) = (n-x(p_n)-y(p_n)) - (n-x(q_n)-y(q_n)) = -(d_x(\phi) + d_y(\phi)),
\end{equation}
so this is not really necessary. See \cref{fig:3d_AO} for an example.

For each of the four models, we define a partition function
\begin{equation}
X_n(c,u,v) = \sum_{\substack{\phi\in\mathcal X \\ |\phi| = n}} c^{v(\phi)} u^{d_x(\phi)} v^{d_y(\phi)}
\end{equation}
and generating function
\begin{equation}
P_X(t;c,u,v) \equiv P(u,v) = \sum_n X_n(c,u,v)t^n = \sum_{\phi\in\mathcal X}t^{|\phi|} c^{v(\phi)} u^{d_x(\phi)} v^{d_y(\phi)}.
\end{equation}

Note that, since $d_x$ and $d_y$ can be negative, $X_n(c,u,v) \in \mathbb Z[c,u,v,\olu,\olv]$.

\subsection{Symmetric and friendly pairs of paths}

\subsubsection{The generating function with zipping only}

Having not yet defined $d(\phi)$, we will first only consider the model with a weight $c$ associated with shared vertices (but no pulling force).

Pairs of paths are grown iteratively in the same way as for two dimensions. A pair of paths is either a single vertex, or can be constructed by appending a new step to each path. Each path has three choices: $+x$, $+y$ or $+z$. When both paths step in the same direction, neither $d_x$ nor $d_y$ change. When one of paths steps $+z$ and the other steps $+x$ (resp.~$+y$), only $d_x$ (resp.~$d_y$) changes. When neither path steps $+z$, both $d_x$ and $d_y$ change. And when the paths step to the same vertex, the pair gains a factor of $c$.

For brevity, write $W^{(i,j)}_{SF} = [u^iv^j]\WSF(u,v)$. Then the above can be encoded with the functional equation
\begin{multline}
\WSF(u,v) = 1 + t(3+u+v+\ol{u}+\ol{v}+u\ol{v}+\ol{u}v)\WSF(u,v) + 3t(c-1)W^{(0,0)}_{SF} \\
+ t(c-1)W^{(1,0)}_{SF} + t(c-1)W^{(0,1)}_{SF} + t(c-1)W^{(-1,0)}_{SF} \\
+ t(c-1)W^{(0,-1)}_{SF} + t(c-1)W^{(-1,1)}_{SF} + t(c-1)W^{(1,-1)}_{SF}.
\end{multline}

It may seem that there are too many unknowns to handle here, but the model has many symmetries we can exploit. If $\phi=(p,q)$ is a pair of paths, define $S_{xy}(\phi)$ to be the pair obtained by replacing every $+x$ step with a $+y$ step, and vice versa, in $p$ and $q$. Similarly define $S_{xz}(\phi)$ and $S_{yz}(\phi)$. Note that, since $d_x(\phi) = d_y(S_{xy}(\phi))$ and $d_y(\phi) = d_x(S_{xy}(\phi))$, if $\phi$ has a shared vertex at step $i$ then so too does $S_{xy}(\phi)$. Similar arguments apply to $S_{xz}(\phi)$ and $S_{yz}(\phi)$. It follows that $v(\phi) = v(S_{xy}(\phi)) = v(S_{xz}(\phi)) = v(S_{yz}(\phi))$, and hence
\begin{equation}
W^{(1,0)}_{SF} = W^{(0,1)}_{SF} = W^{(-1,0)}_{SF} = W^{(0,-1)}_{SF} = W^{(-1,1)}_{SF} = W^{(1,-1)}_{SF}.
\end{equation}

Thus
\begin{equation}\label{eqn:WSF_fe_simpler}
\WSF(u,v) = 1 + t(3+u+v+\ol{u}+\ol{v}+u\ol{v}+\ol{u}v)\WSF(u,v) + 3t(c-1)\WSF^{(0,0)} + 6t(c-1)\WSF^{(1,0)}.
\end{equation}

Next, by considering only those pairs which end at a shared vertex,
\begin{equation}
\WSF^{(0,0)} = 1+3tc\WSF^{(0,0)} + 6tc\WSF^{(1,0)}.
\end{equation}
We then arrive at
\begin{equation}\label{eqn:3d_SF_eqn}
\WSF(u,v) = \frac1c + t(3+u+v+\ol{u}+\ol{v}+u\ol{v}+\ol{u}v)\WSF(u,v) + \left(1-\frac1c\right)\WSF^{(0,0)}.
\end{equation}

Write~\eqref{eqn:3d_SF_eqn} in kernel form
\begin{equation}\label{eqn:3d_SF_kernelised}
K(u,v)\WSF(u,v) = \frac1c + \left(1-\frac1c\right)\WSF^{(0,0)}
\end{equation}
where $K(u,v) = 1-t(3+u+v+\ol{u}+\ol{v}+u\ol{v}+\ol{u}v)$.

There are obvious similarities between~\eqref{eqn:3d_SF_kernelised} and~\eqref{eqn:PAF_cancelled}. However, the key difference here is that the coefficients of $\WSF(u,v)$ are \emph{Laurent polynomials} in $u$ and $v$. So while $K(u,v)$ does have a root in $u$ (or, by symmetry, $v$) which is a power series in $t$, it cannot be validly substituted into~\eqref{eqn:3d_SF_kernelised}.

However, we have another way to approach this problem. Rearranging,
\begin{equation}\label{eqn:SF_kernel_rearrange}
\WSF(u,v) = \frac{1}{cK(u,v)} + \frac{c-1}{cK(u,v)}\WSF^{(0,0)}.
\end{equation}
Let
\begin{align}
X(t) \equiv X &= [u^0v^0]\frac{1}{K(u,v)} \\
&= 1 + 3t + 15t^2 + 93t^3 + 639t^4 + 4653t^5 + \dots \\
&= \sum_{n=0}^\infty x_n t^n
\end{align}
where
\begin{equation}\label{eqn:xc_binoms}
x_n = \sum_{k=0}^n \binom{2k}{k}\binom{n}{k}^2
\end{equation}
(OEIS sequence A002893). The series $X$ can be written in terms of $K$, the complete elliptic integral of the first kind:
\begin{equation}
X = \frac{2\sqrt{2}}{\pi\sqrt{1-6t-3t^2+\sqrt{(1-t)^3(1-9t)}}}K\left(\frac{8t^{\frac32}}{1-6t-3t^2+\sqrt{(1-t)^3(1-9t)}}\right)
\end{equation}

Then by extracting the constant term with respect to $u$ and $v$ from~\eqref{eqn:SF_kernel_rearrange}, we find
\begin{equation}
\WSF^{(0,0)} = \frac{X}{c+(1-c)X}
\end{equation}
the generating function of asymmetric and friendly pairs of paths which start and end together.

\subsubsection{Incorporating the unzipping force}

We can model zipping with this, but it does not allow us to model a force pulling on the two ends. For that, we substitute back into~\eqref{eqn:SF_kernel_rearrange}:
\begin{equation}\label{eqn:SF_uv_full_eqn}
\WSF(u,v) = \frac{1}{cK(u,v)} + \frac{(c-1)X}{cK(u,v)(c+(1-c)X)} = \frac{1}{K(u,v)(c+(1-c)X)}.
\end{equation}

With two directed paths in two dimensions, the statistic $d(\phi)$ was useful not only for modelling a force applied at the endpoints, but also played a part in the solutions to the functional equations, with the variable $y$ temporarily serving as a ``catalytic variable''. Here, the catalytic variables are $u$ and $v$, but $d_x$ and $d_y$ are not exactly what we need in order to incorporate the force.

Let $d^*(\phi)$ be the Euclidean distance between the endpoints of the two paths. We have
\begin{equation}\label{eqn:3d_d*}
d^*(\phi) = \sqrt{2\left(d_x(\phi)^2 + d_x(\phi)d_y(\phi) + d_y(\phi)^2\right)}.
\end{equation}
Since this is not a linear function of $d_x$ and $d_y$, there is no simple evaluation of $u$ and $v$ which allows us to introduce a Boltzmann weight of the form $y^{d^*(\phi)}$.

Instead, let $d(\phi)$ be the minimum number of steps required for the two paths $p$ and $q$ to reach a shared vertex. This can be written as a piecewise linear function of $d_x$ and $d_y$:
\begin{equation}\label{eqn:ij_num_steps_together}
d(\phi) = \begin{cases} d_x(\phi)+d_y(\phi) & \text{ if } d_x(\phi),d_y(\phi)\geq 0 \\
-d_x(\phi)-d_y(\phi) & \text{ if } d_x(\phi),d_y(\phi) \leq 0 \\
d_x(\phi) & \text{ if } 0 \leq -d_y(\phi) \leq d_x(\phi) \\
-d_y(\phi) & \text{ if } 0 \leq d_x(\phi) \leq -d_y(\phi) \\
d_y(\phi) & \text{ if } 0 \leq -d_x(\phi) \leq d_y(\phi) \\
-d_x(\phi) & \text{ if } 0 \leq d_y(\phi) \leq -d_x(\phi)
\end{cases}
\end{equation}

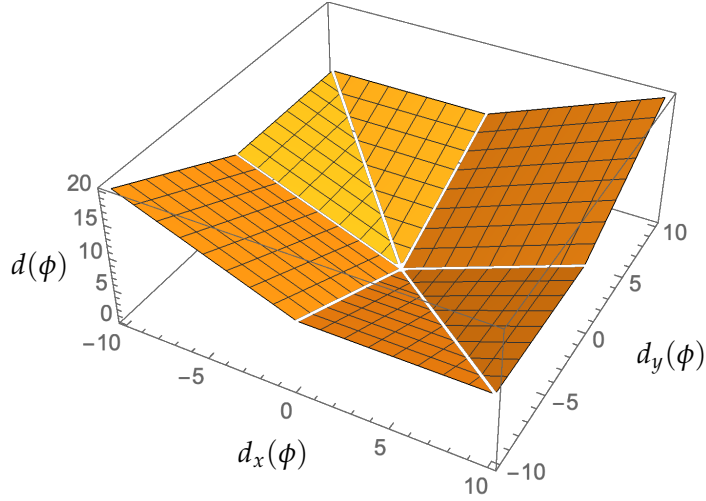
\begin{figure}
    \centering
    \begin{tikzpicture}
    \node at (0,0) {\includegraphics[width=0.5\textwidth]{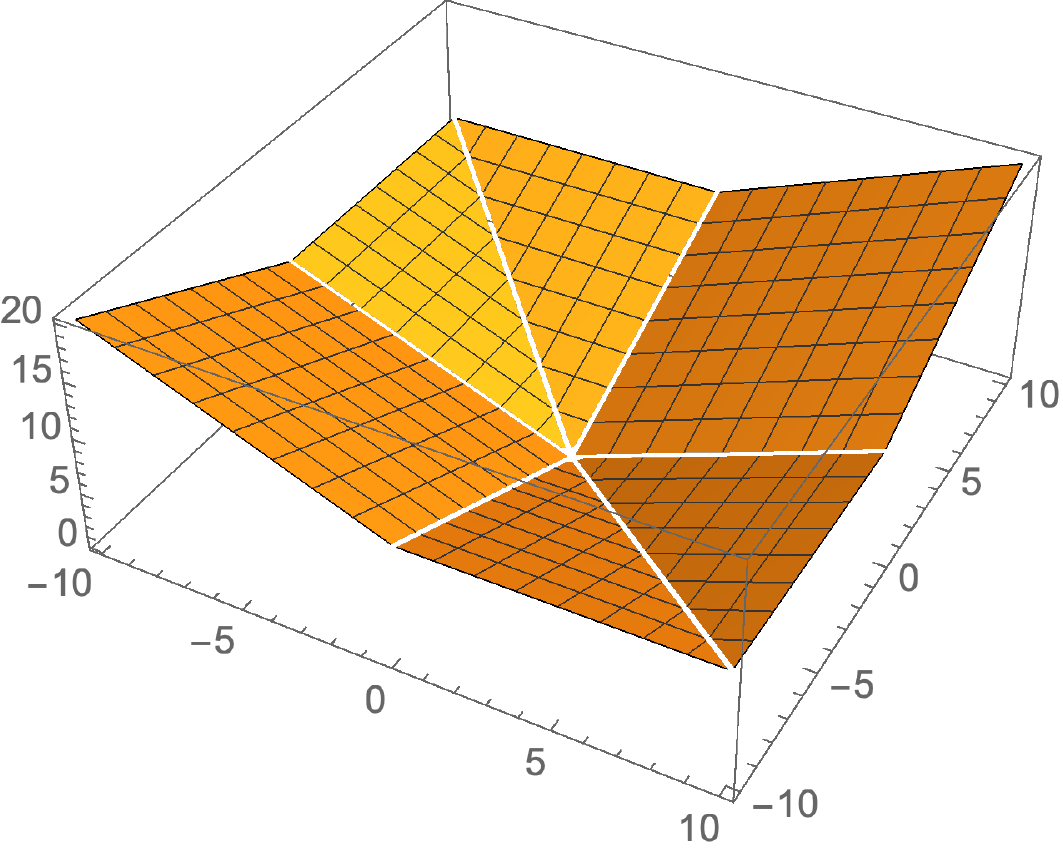}};
    \node at (-1.4,-2.7) {$d_x(\phi)$};
    \node at (3.9,-1.4) {$d_y(\phi)$};
    \node at (-4.5,-0.2) {$d(\phi)$};
    \end{tikzpicture}
    \caption{A plot of $d(\phi)$ versus $d_x(\phi)$ and $d_y(\phi)$.}
    \label{fig:dphi}
\end{figure}

See \cref{fig:dphi}. This looks complicated, but we can again exploit the inherent symmetries of the model. By applying all possible combinations of the maps $S_{xz}$ and $S_{yz}$ (note that $S_{xy}$ is not necessary -- this will be important later when we come to the asymmetric models), we have
\begin{equation}\label{eqn:SF_symmetries}
\WSF(u,v) = \WSF(\ol{u}v,v) = \WSF(\ol{v},u\ol{v}) = \WSF(\ol{v},\ol{u}) = \WSF(\ol{u}v,\ol{u}) = \WSF(u,u\ol{v}).
\end{equation}
If a configuration has weight $u^{d_x(\phi)}v^{d_y(\phi)}$, with $d_x(\phi)$ and $d_y(\phi)$ satisfying one of the six conditions in~\eqref{eqn:ij_num_steps_together}, then~\eqref{eqn:SF_symmetries} implies that it can be uniquely mapped to a configuration which satisfies any one of the other five conditions. In other words, there is a six-fold symmetry. To study the phase diagram it thus suffices to focus on only one of the six symmetry classes. We will focus on the case $d_x(\phi),d_y(\phi)\geq 0$.

Before proceeding with the solution, we are now also able to compare $d^*(\phi)$ and $d(\phi)$. With $d_x(\phi),d_y(\phi)\geq 0$ and manipulating~\eqref{eqn:3d_d*},
\begin{align}
&&\sqrt{d_x(\phi)^2 + 2d_x(\phi)d_y(\phi) + d_y(\phi)^2} &\leq d^*(\phi) \leq \sqrt{2\left(d_x(\phi)^2 + 2d_x(\phi)d_y(\phi) + d_y(\phi)^2\right)} \\ 
&\iff& d_x(\phi) + d_y(\phi) &\leq d^*(\phi) \leq \sqrt{2}\left(d_x(\phi) + d_y(\phi)\right) \\
&\iff& d(\phi) &\leq d^*(\phi) \leq \sqrt{2}d(\phi) \\
&\iff& {\textstyle\frac{1}{\sqrt{2}}d^*(\phi)} &\leq \; d(\phi)\, \leq d^*(\phi).
\end{align}
We thus see that, while $d(\phi)$ does not exactly correspond to the distance between the two endpoints, it is bounded above and below by constant multiples of this distance.

Let $\WSF^{++}(u,v)$ be the part of $\WSF(u,v)$ with non-negative powers in $u$ and $v$. Then from~\eqref{eqn:SF_uv_full_eqn},
\begin{equation}
\WSF^{++}(u,v) = \frac{X^{++}(u,v)}{c+(1-c)X}
\end{equation}
where $X^{++}(u,v)$ is the part of $\frac{1}{K(u,v)}$ with non-negative powers in $u$ and $v$. We wish to assign weight $y^{d(\phi)}$, so let 
\begin{align}
X^*(t;y) \equiv X^*(y) &= X^{++}(u,v)|_{u=v=y} \\
&= 1 + t(3+2y) + t^2(15+16y+4y^2) + t^3(93 + 120 y + 54 y^2 + 8 y^3) + \dots
\end{align}
and define
\begin{equation}
\WSF^*(t;c,y) = \frac{X^*(y)}{c+(1-c)X}
\end{equation}
as the generating function we are interested in.

\subsubsection{The dominant singularity}

Now there are three singularities of interest here: the dominant singularity of $X$, the dominant root of $c+(1-c)X$, and the dominant singularity of $X^*(y)$. As for $X$, one finds that
\begin{equation}\label{eqn:X_near_sing}
X \underset{t\to{\frac19}^-}{\sim} A \log\left(\frac{1}{1-9t}\right) + B + o(1), \qquad \text{ where } A = \frac{3\sqrt{3}}{4\pi} \text{ and } B = (3\log2)A.
\end{equation}
Thus
\begin{equation}
x_n \sim A\cdot\frac{9^n}{n}.
\end{equation}

So $X$ has radius of convergence $\frac19$, and diverges as $t\to\frac19$ from below. It follows that if $0\leq c <1$ then $c+(1-c)X$ has no roots for $t\in[0,\frac19]$. If $c>1$ then there will be a root, say $\rho_{SF}(c)$, smaller than $\frac19$.

Unfortunately the complicated nature of $X$ means that we have no way of getting an explicit expression for $\rho_{SF}(c)$. We can, however, determine its behaviour as $c\to1^+$ and as $c\to\infty$. For small $c$, some numerical investigation shows that
\begin{equation}\label{eqn:rhoSF_cto1}
\rhoSF(c) \underset{c\to1^+}{=} \frac19 - \exp\left(\displaystyle \frac{\alpha}{c-1} + \beta + O(c-1)\right).
\end{equation}
Substituting this into~\eqref{eqn:X_near_sing} and solving $X = \frac{c}{c-1}$, we find
\begin{equation}
\alpha = -\frac{4\pi}{3\sqrt{3}} \qquad \text{and}\qquad \beta = -\frac{4\pi}{3\sqrt{3}} - \log\left(\frac{9}{8}\right).
\end{equation}

For large $c$, we can use Lagrange inversion and~\eqref{eqn:xc_binoms} to obtain an asymptotic expansion:
\begin{equation}\label{eqn:rhosf_lagrange}
\rhoSF(c) \underset{c\to\infty}{=} \olc - 2\olc^2 - 2\olc^3 + 8\olc^5 + O\big(\olc^{6}\big),
\end{equation}
where $\olc = \frac{1}{3c}$. The expression~\eqref{eqn:rhosf_lagrange} becomes more accurate with increasing $c$.

Next we turn to $X^*(y)$. This is a D-finite function but we do not have a simple explicit expression. However, we can compute the asymptotic behaviour of the coefficients (and thus the dominant singularity). Observe that
\begin{align}
[t^n]\frac{1}{K(u,v)} &= (3+u+v+\ol{u}+\ol{v}+u\ol{v}+\ol{u}v)^n \\
&= \sum_{k_1+\dots+k_7 = n} \binom{n}{k_1,\dots,k_7} 3^{k_1} u^{k_2-k_4+k_6-k_7} v^{k_3-k_5-k_6+k_7}.
\end{align}
Then
\begin{equation}
[t^n]X^{*}(y) = \sum_{\substack{k_1+\dots+k_7 = n \\ -k_2+k_4 \leq k_6-k_7 \leq k_3-k_5}} \binom{n}{k_1,\dots,k_7} 3^{k_1} y^{k_2+k_3-k_4-k_5}.
\end{equation}
This six-fold sum can be written out over the ranges
\begin{align}
k_2 &= 0,\dots,n \\
k_3 &= 0,\dots,n-k_2 \\
k_4 &= 0,\dots,n-k_2-k_3 \\
k_5 &= 0,\dots,n-k_2-k_3-k_4 \\
k_6 &= 0,\dots,n-k_2-k_3-k_4-k_5 \\
k_7 &= \max\{0,-k_3+k_5+k_6\},\dots,\min\{n-k_2-k_3-k_4-k_5-k_6,k_2-k_4+k_6\} \label{eqn:k7_range}\\
k_1 &= n-k_2-k_3-k_4-k_5-k_6-k_7.
\end{align}
To compute the asymptotics we replace the sums with integrals and apply Stirling's approximation:
\[n! \sim \sqrt{2\pi n}\left(\frac{n}{e}\right)^n\left(1+O\left(\frac1n\right)\right).\]
When $y>1$, the sum (integral) is dominated by terms with $k_1,\dots,k_7$ all $O(n)$. We thus set $k_i = \kappa_i n$ for constants $\kappa_i$ (to be determined), where $\kappa_1 = 1-\kappa_2-\dots-\kappa_7$. The dominant term in the integrand is then
\begin{equation}
I_n(y) = \frac{3^{\kappa_1 n}y^{(\kappa_2+\kappa_3-\kappa_4-\kappa_5)n}}{8n^3\pi^3}\prod_{i=1}^7 \kappa_i^{-(\kappa_i n+1/2)}.\label{eqn:integrand_In}
\end{equation}
We wish to find the values of the $\kappa_i$ which maximise this, or rather, its growth rate. To do this, we take $\frac1n \log I_n$, take the derivative with respect to $\kappa_i$ for $i=2,\dots,6$ (separately), and then take the limit $n\to\infty$ in each. This gives the six terms
\begin{align}
&\log y-\log 3 -\log\kappa_2 + \log\kappa_1 \\
&\log y-\log 3 -\log\kappa_3 + \log\kappa_1 \\
&-\log y-\log 3 -\log\kappa_4 + \log\kappa_1 \\
&-\log y-\log 3 -\log\kappa_5 + \log\kappa_1 \\
&-\log 3 -\log\kappa_4 + \log\kappa_1 \\
&-\log 3 -\log\kappa_5 + \log\kappa_1
\end{align}
To maximise we set all of these to 0, and arrive at the solutions
\begin{align}
\kappa_1 &= \frac{3y}{(2+y)(1+2y)} \\
\kappa_2 = \kappa_3 &= \frac{y^2}{(2+y)(1+2y)} \\
\kappa_4 = \kappa_5 &= \frac{1}{(2+y)(1+2y)} \\
\kappa_6 = \kappa_7 &= \frac{y}{(2+y)(1+2y)}
\end{align}
(Note that for $y>1$, the condition~\eqref{eqn:k7_range} is automatically satisfied. That is, when $y>1$ it is the first of the six conditions in~\eqref{eqn:ij_num_steps_together} which dominates.)

Upon substitution back into~\eqref{eqn:integrand_In}, we find that the growth rate of $I_n(y)$, and hence of $[t^n]X^{*}(y)$, is $\frac{(2+y)(1+2y)}{y}$. (We could actually compute the integral to find the full asymptotics of $[t^n]X^{*}(y)$, but this is not necessary to get the free energy.) For $y>1$, the critical point of $X^*(y)$ is thus $\sigSF(y) = \frac{y}{(2+y)(1+2y)}$.

Putting all this together and determining which singularities dominate where, we find
\begin{align}
\tSF(c,y) &= \min\{\textstyle\frac19, \rhoSF(c), \sigSF(y)\} \\
&= \begin{cases} \textstyle\frac19 & \text{ if } c\leq 1 \text{ and } y\leq 1 \\ 
\sigSF(y) & \text{ if } y \geq \max\{1,f(\rhoSF(c))\} \\
\rhoSF(c) & \text{ if } c \geq 1 \text{ and } y \leq f(\rhoSF(c))
\end{cases}
\end{align}
where
\begin{equation}\label{eqn:tSF_f_defn}
f(x) = \frac{1-5x + \sqrt{(1-x)(1-9x)}}{4x}.
\end{equation}

\subsection{Symmetric and osculating pairs of paths}
\subsubsection{The generating function}

It is straightforward to repeat the above procedure for osculating paths. Let $\WSO(u,v)$ be the analogue of $\WSF(u,v)$. Then the equivalent of~\eqref{eqn:WSF_fe_simpler} is 
\begin{equation}\label{eqn:WSO_fe_simpler}
\WSO(u,v) = 1 + t(3+u+v+\ol{u}+\ol{v}+u\ol{v}+\ol{u}v)\WSO(u,v) + -3t\WSO^{(0,0)} + 6t(c-1)\WSO^{(1,0)},
\end{equation}
the difference being that the walks cannot step in the same directions when $d_x=d_y=0$. We then have
\begin{equation}
\WSO^{(0,0)} = 1 + 6tc\WSO^{(1,0)}.
\end{equation}
Substituting,
\begin{equation}
K(u,v)\WSO(u,v) = \frac{1}{c} + \left(1-\frac1c -3t\right)\WSO^{(0,0)},
\end{equation}
with $K(u,v)$ as defined in the previous section. Also using the same $X$ as before, we find
\begin{equation}
\WSO(u,v) = \frac{1+6tX}{K(u,v)\left(c+(1-c+3tc)X\right)}.
\end{equation}
To incorporate the unzipping force, we can again make use of the symmetries of the model, and only consider the cases with $d_x,d_y\geq0$. So we focus on
\begin{equation}\label{eqn:WSO*_form}
\WSO^*(t;c,y) = \frac{X^*(y)(1+6tX)}{c+(1-c+3tc)X}.
\end{equation}

\subsubsection{The dominant singularity}

This time the denominator of~\eqref{eqn:WSO*_form} has a root $t=\rhoSO(c)$ when $c>\frac32$. As $c\to\frac{3}{2}^+$ we observe similar behaviour to~\eqref{eqn:rhoSF_cto1}:
\begin{equation}\label{eqn:rhoSO_cto32}
\rhoSO(c) \underset{c\to\frac{3}{2}^+}{=} \frac19 - \exp\left(\displaystyle \frac{\alpha}{c-\textstyle\frac32} + \beta + O(c-\textstyle\frac32)\right).
\end{equation}
Again using~\eqref{eqn:X_near_sing} and solving $c+(1-c+3tc)X=0$, we find
\begin{equation}
\alpha = -\sqrt{3}\pi \qquad\text{and}\qquad \beta = -\frac{2\pi}{\sqrt{3}}-\log\left(\frac98\right).
\end{equation}

As $c\to\infty$, we can again use Lagrange inversion to determine the behaviour of $\rhoSO(c)$. This time, letting $\tilde{c} = \frac{1}{4\sqrt{6c}}$, we have
\begin{equation}
\rhoSO(c) \underset{c\to\infty}{=} 4\tilde{c} - 40\tilde{c}^2 + 40\tilde{c}^3 + 256\tilde{c}^4 + 1336\tilde{c}^5 + O(\tilde{c}^6).
\end{equation}

The $y$-dependence of $\WSO^*$ and $\WSF^*$ is the same, i.e.~the factor $X^*(y)$. So let $\sigSO(y) = \sigSF(y)$.

Then 
\begin{align}
\tSO(c,y) &= \min\{\textstyle\frac19, \rhoSO(c), \sigSO(y)\} \\
&= \begin{cases} \textstyle\frac19 & \text{ if } c\leq \frac32 \text{ and } y\leq 1 \\ 
\sigSF(y) & \text{ if } y \geq \max\{1,f(\rhoSO(c))\} \\
\rhoSF(c) & \text{ if } c \geq \frac32 \text{ and } y \leq f(\rhoSO(c))
\end{cases}
\end{align}
where $f$ is as defined in~\eqref{eqn:tSF_f_defn}.

\subsection{Asymmetric and friendly pairs of paths}

\subsubsection{The generating function}

Things are a little more complicated here, as $S_{xy}$ is no longer a valid symmetry of the model. The main functional equation is
\begin{multline}
\WAF(u,v) = 1 + t(3+u+v+\ol{u}+\ol{v}+u\ol{v}+\ol{u}v)\WAF(u,v) + 3t(c-1)\WAF^{(0,0)} \\ - t(u+\ol{u}v+\ol{v})\WAF^{(0,0)} + 3t(c-1)\WAF^{(1,0)} + 3t(c-1)\WAF^{(0,1)}.
\end{multline}
Then using
\begin{equation}
\WAF^{(0,0)} = 1 + 3tc\WAF^{(0,0)} + 3tc\WAF^{(1,0)} + 3tc\WAF^{(0,1)},
\end{equation}
we arrive at
\begin{equation}
K(u,v)\WAF(u,v) = \frac1c + \left(1-\frac1c-t(u+\ol{u}v+\ol{v})\right)\WAF^{(0,0)}
\end{equation}
or alternatively
\begin{equation}\label{eqn:WAF_kernel_form}
\WAF(u,v) = \frac{1}{cK(u,v)} + \frac{c-1-tc(u+\ol{u}v+\ol{v})}{cK(u,v)}\WAF^{(0,0)}.
\end{equation}

Now let
\begin{equation}
Y(t) \equiv Y = [u^{-1}v^0]\frac{1}{K(u,v)} = [u^{1}v^{-1}]\frac{1}{K(u,v)} = [u^0v^1]\frac{1}{K(u,v)}.
\end{equation}
Extracting the coefficient of $u^0v^0$ in~\eqref{eqn:WAF_kernel_form}, 
\begin{equation}\label{eqn:WAF_u0v0}
\WAF^{(0,0)} = \frac{X}{c} + \frac{(c-1)X}{c} \WAF^{(0,0)} - 3tY\WAF^{(0,0)}.
\end{equation}
But now we also have
\begin{equation}
Y = [u^{1}v^0]\frac{1}{K(u,v)} = [u^{-1}v^{1}]\frac{1}{K(u,v)} = [u^0v^{1}]\frac{1}{K(u,v)},
\end{equation}
so $X=1+3tX + 6tY$. Substituting into~\eqref{eqn:WAF_u0v0} and solving,
\begin{equation}
\WAF^{(0,0)} = \frac{2X}{c+(2-c-3tc)X}.
\end{equation}
Then
\begin{equation}
\WAF(u,v) = \frac{1}{K(u,v)(c+(2-c-3tc)X)}\left[1+(1-3t)X-2t(u+\ol{u}v+\ol{v})X\right]
\end{equation}

Now $\WAF$ still satisfies the same six-fold symmetry as $\WSF$ and $\WSO$ as per~\eqref{eqn:SF_symmetries} (because $S_{xy}$ was not required there), so we can still incorporate the unzipping force by restricting to those configurations with $d_x,d_y\geq0$ and setting $u=v=y$. Define
\begin{align}
X^\rightarrow(u,v) &= tu[u^{\geq-1}v^{\geq0}]\frac{1}{K(u,v)}\\ X^\nwarrow(u,v) &= t\ol{u}v[u^{\geq1}v^{\geq-1}]\frac{1}{K(u,v)} \\ X^\downarrow(u,v) &= t\ol{v}[u^{\geq0}v^{\geq1}]\frac{1}{K(u,v)}.
\end{align}
Then
\begin{multline}
\WAF^{++}(u,v) = \frac{1+(1-3t)X}{c+(2-c-3tc)X}X^{++}(u,v) \\ - \frac{2X}{c+(2-c-3tc)X}\left[X^\rightarrow(u,v) + X^\nwarrow(u,v) + X^\downarrow(u,v)\right].
\end{multline}
Finally, let
\begin{equation}
X^\dagger(y) = \left.X^\rightarrow(u,v) + X^\nwarrow(u,v) + X^\downarrow(u,v)\right|_{u=v=y}.
\end{equation}
Then
\begin{equation}\label{eqn:WAF_tcy}
\WAF^*(t;c,y) = \frac{(1+(1-3t)X)X^*(y)-2XX^\dagger(y)}{c+(2-c-3tc)X}.
\end{equation}

\subsubsection{The dominant singularity}

The critical value of $c$ is again $\frac32$, with the denominator having a root $\rhoAF(c)$ if $c>\frac32$. For small $c$ we have
\begin{equation}\label{eqn:rhoAF_cto32}
\rhoAF(c) \underset{c\to\frac{3}{2}^+}{=} \frac19 - \exp\left(\displaystyle \frac{\alpha}{c-\textstyle\frac32} + \beta + O(c-\textstyle\frac32)\right).
\end{equation}
We determine $\alpha$ and $\beta$ by substituting~\eqref{eqn:X_near_sing} into the denominator of~\eqref{eqn:WAF_tcy} and taking the limit $c\to\frac32$. In this case,
\begin{equation}
\alpha = -\frac{\sqrt{3}\pi}{2} \qquad\text{and}\qquad \beta = -\frac{\pi}{\sqrt{3}}-\log\left(\frac98\right).
\end{equation}
Meanwhile, as $c\to\infty$, we have
\begin{equation}
\rhoAF(c) \underset{c\to\infty}{=} \ol{c} - \ol{c}^2 - 3\ol{c}^3 - 10\ol{c}^4 - 34\ol{c}^5 + O(\ol{c}^6)
\end{equation}
where $\ol{c} = \frac{1}{3c}$ as before.

As for the $y$-dependence, we now have the function $X^\dagger(y)$ in addition to $X^*(y)$. However, note that $X^{++}(u,v)$ counts configurations of symmetric and friendly paths (with no $c$ weight) with $d_x,d_y\geq0$, while $X^\rightarrow(u,v) + X^\nwarrow(u,v) + X^\downarrow(u,v)$ counts a \emph{subset} of those paths -- namely those ending with a $(+x,+z)$, $(+y,+x)$ or $(+z,+y)$ pair of steps. Hence, considered as formal power series with non-negative coefficients, $X^\dagger(y) \leq X^*(y)$, and so the dominant singularity of $X^\dagger(y)$ is bounded below by that of $X^*(y)$. So nothing new happens here, and we can set $\sigAF(y) = \sigSF(y)$.

Then
\begin{align}
\tAF(c,y) &= \min\{\textstyle\frac19, \rhoAF(c), \sigAF(y)\} \\
&= \begin{cases} \textstyle\frac19 & \text{ if } c\leq \frac32 \text{ and } y\leq 1 \\ 
\sigAF(y) & \text{ if } y \geq \max\{1,f(\rhoAF(c))\} \\
\rhoAF(c) & \text{ if } c \geq \frac32 \text{ and } y \leq f(\rhoAF(c))
\end{cases}
\end{align}
where $f$ is as defined in~\eqref{eqn:tSF_f_defn}.

\subsection{Asymmetric and osculating pairs of paths}
\subsubsection{The generating function}

The last model we consider has both the asymmetric and osculating restrictions. The main functional equation is
\begin{multline}
\WAO(u,v) = 1 + t(3+u+v+\ol{u}+\ol{v}+u\ol{v}+\ol{u}v)\WAO(u,v) \\ - t(3+u+\ol{u}v+\ol{v})\WAO^{(0,0)} + 3t(c-1)\WAO^{(1,0)} + 3t(c-1)\WAO^{(0,1)}.
\end{multline}
Using
\begin{equation}
\WAO^{(0,0)} = 1 + 3tc\WAO^{(1,0)} + 3tc\WAO^{(0,1)}
\end{equation}
we get
\begin{equation}
K(u,v)\WAO(u,v) = \frac1c + \left(1-\frac1c-t(3+u+\ol{u}v+\ol{v})\right)\WAO^{(0,0)}.
\end{equation}
Then
\begin{equation}
\WAO^{(0,0)} = \frac{X}{c} + \frac{(c-1-3tc)X}{c}\WAO^{(0,0)} -3tY\WAO^{(0,0)},
\end{equation}
and using $X=1+3tX+6tY$, we have
\begin{equation}
\WAO^{(0,0)} = \frac{2X}{c+(2-c+3tc)X}.
\end{equation}
Then
\begin{equation}
\WAO(u,v) = \frac{1}{K(u,v)(c+(2-c+3tc)X)}\left[1 + (1-3t)X -2t(u+\ol{u}v+\ol{v})X\right].
\end{equation}
Using the same technique as for $\WAF$,
\begin{multline}
\WAO^{++}(u,v) = \frac{1+(1-3t)X}{c+(2-c+3tc)X}X^{++}(u,v) \\ - \frac{2X}{c+(2-c-3tc)X}\left[X^\rightarrow(u,v) + X^\nwarrow(u,v) + X^\downarrow(u,v)\right],
\end{multline}
and so finally
\begin{equation}\label{eqn:WAO_tcy}
\WAO^*(t;c,y) = \frac{(1+(1-3t)X)X^*(y)-2XX^\dagger(y)}{c+(2-c+3tc)X}.
\end{equation}

\subsubsection{The dominant singularity}

This time the critical value of $c$ is 3, with the denominator having a root $\rhoAO(c)$ if $c>3$. As $c\to3^+$, we have
\begin{equation}\label{eqn:rhoAO_cto3}
\rhoAO(c) \underset{c\to3^+}{=} \frac19 - \exp\left(\displaystyle \frac{\alpha}{c-3} + \beta + O(c-3)\right)
\end{equation}
with
\begin{equation}
\alpha = -2\sqrt{3}\pi \qquad\text{and}\qquad \beta = -\frac{2\pi}{\sqrt{3}}-\log\left(\frac98\right).
\end{equation}
As $c\to\infty$,
\begin{equation}
\rhoAO(c) \underset{c\to\infty}{=} 4\hat{c} - 40\hat{c}^2 + 40\hat{c}^3 + 256\hat{c}^4 + 1336\hat{c}^5 + O(\hat{c}^6)
\end{equation}
where $\hat{c} = \frac{1}{4\sqrt{3c}}$.

As with the three earlier cases, the $y$-dependence comes from $X^*(y)$, so set $\sigAO(y) = \sigSF(y)$.

Then
\begin{align}
\tAO(c,y) &= \min\{\textstyle\frac19, \rhoAO(c), \sigAO(y)\} \\
&= \begin{cases} \textstyle\frac19 & \text{ if } c\leq 3 \text{ and } y\leq 1 \\ 
\sigAO(y) & \text{ if } y \geq \max\{1,f(\rhoAO(c))\} \\
\rhoAO(c) & \text{ if } c \geq 3 \text{ and } y \leq f(\rhoAO(c))
\end{cases}
\end{align}

\subsection{Phase diagrams}

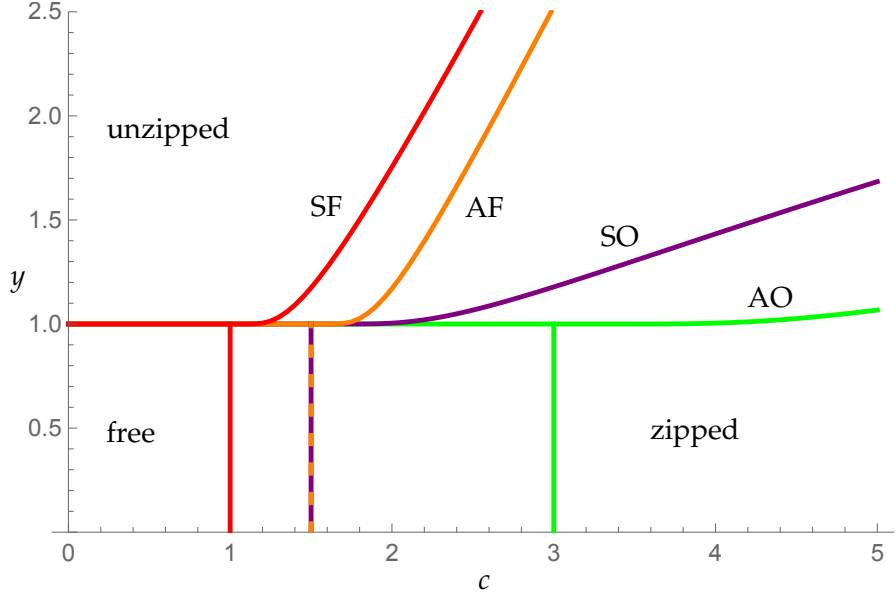
\begin{figure}
\centering
\begin{tikzpicture}
\node at (0,0) [anchor=south west] {\includegraphics[width=0.7\textwidth]{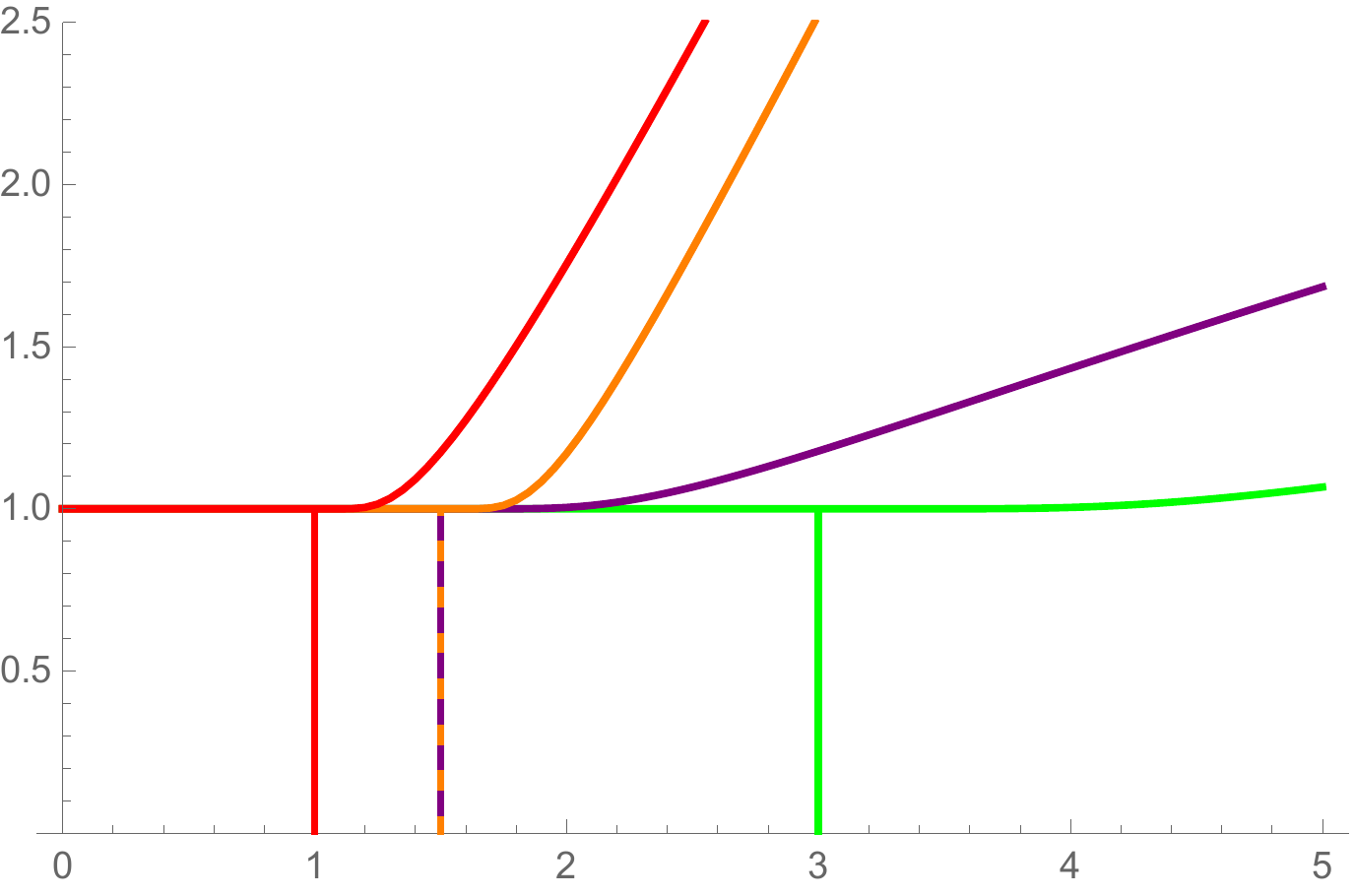}};
\node at (4.1,5) {SF};
\node at (6.2,5) {AF};
\node at (8,4.6) {SO};
\node at (10,3.8) {AO};
\node at (6.2,0) {$c$};
\node at (0,4) {$y$};
\node at (9,2) {zipped};
\node at (2,6) {unzipped};
\node at (1.5,2) {free};
\end{tikzpicture}
\caption{The phase boundaries for the four three-dimensional models. The vertical lines are the boundaries between the free and zipped phases; the horizontal lines are the boundaries between free and unzipped; and the sloping curves are the unzipped-zipped boundaries. The free-zipped boundaries for the AF and SO models coincide.}
\label{fig:3D_phasediagrams}
\end{figure}

We plot the four phase diagrams together in \cref{fig:3D_phasediagrams}.

For fixed $y\leq1$, the zipping transitions occur with increasing $c$ in the order
\[\mathrm{SF} < \mathrm{AF} \equiv \mathrm{SO} < \mathrm{AO}.\]
This can be understood in the same way as the two-dimensional case: SF loses no entropy at a contact, AF and SO each lose three of the nine step choices, and AO loses six choices.

For fixed $y>1$, the unzipped-zipped transitions occur with increasing $c$ in the order
\[\mathrm{SF} < \mathrm{AF} < \mathrm{SO} < \mathrm{AO}.\]
The fact that the AF model ``zips'' together before the SO model can be understood by observing that the zipped phase for the AF model has twice the density of contacts of the SO model, and so $\rhoAF(c)$ decreases more quickly (with increasing $c$) than $\rhoSO(c)$.

In all cases the free-zipped and free-unzipped phase transitions are second-order, while the unzipped-zipped transitions are first-order.

\section{Conclusion}\label{sec:conclusion}

We have defined and analysed four different models of interacting pairs of directed polymers, in two and three dimensions. The different models are classified according to whether the polymers are able to share edges or only sites, and according to the allowed symmetries between the pair. In each case we incorporate two Boltzmann weights -- one to control the strength of the attraction/repulsion between the polymers, and another to model a force pulling apart the ends. The models exhibit qualitatively similar but quantitatively different phase diagrams, which have been computed exactly for two dimensions and (partly) numerically for three dimensions.

These models can be enhanced in a number of ways. One would be to include a Boltzmann weight to control the flexibility or stiffness of the polymers; another would be to introduce an impenetrable surface with which the polymers can interact. A further possibility would be to analyse how the polymers twist around one another.

\begingroup\emergencystretch=1em

\printbibliography

\endgroup

\end{document}